\documentclass[runningheads]{llncs}
\usepackage[T1]{fontenc}
\usepackage{graphicx}
\usepackage{booktabs}
\usepackage{multirow}
\usepackage{amsmath}
\usepackage{subcaption}
\usepackage{pgfplots}
\usepackage{cleveref}
\usepackage[dvipsnames]{xcolor}
\usepackage{tikz}
\usetikzlibrary{matrix}
\usetikzlibrary{shapes}
\usetikzlibrary{calc}
\usetikzlibrary{plotmarks}
\pgfplotsset{compat=1.18}
\newcommand{\repeatthanks}{\textsuperscript{\thefootnote}}

\begin{document}
\title{A Reproducible and Fair Evaluation of Partition-aware Collaborative Filtering}
\titlerunning{A Reproducible and Fair Evaluation of Partition-aware CF}
%
\author{Domenico de Gioia\thanks{Corresponding authors: Domenico de Gioia (\email{domenico.degioia@poliba.it}) and Claudio Pomo (\email{claudio.pomo@poliba.it}).}\inst{1}\orcidID{0009-0000-2194-3145} \and
Claudio Pomo\repeatthanks\inst{1}\orcidID{0000-0001-5206-3909} \and\\
Ludovico Boratto\inst{2}\orcidID{0000-0002-6053-3015} \and 
Tommaso Di Noia\inst{1}\orcidID{0000-0002-0939-5462}}

\authorrunning{D. de Gioia et al.}

\institute{Politecnico di Bari, Bari, Italy\\ \email{name.surname@poliba.it} \and
University of Cagliari, Cagliari, Italy\\
\email{ludovico.boratto@acm.org}}
%


\maketitle              
\begin{abstract}
Similarity-based collaborative filtering (CF) models have long demonstrated strong offline performance and conceptual simplicity. However, their scalability is limited by the quadratic cost of maintaining dense item–item similarity matrices. Partitioning-based paradigms have recently emerged as an effective strategy to balance effectiveness and efficiency, allowing models to learn local similarities within coherent subgraphs while maintaining limited global context. In this work, we focus on the Fine-tuning Partition-aware Similarity Refinement (FPSR) framework, a prominent representative of this family, and its extension FPSR+. Reproducible evaluation of partition‑aware collaborative filtering remains challenging, as prior FPSR/FPSR+ reports often rely on splits of unclear provenance and omit some similarity‑based baselines, complicating fair comparison. We present a transparent, fully reproducible benchmark of FPSR and FPSR+.
Based on our results, the family of FPSR models does not consistently perform at the highest level. Overall, it remains competitive, validates its design choices, and shows significant advantages in long-tail scenarios. This highlights the accuracy–coverage trade-offs resulting from partitioning, global components, and hub design. Our investigation clarifies when partition‑aware similarity modeling is most beneficial and offers actionable guidance for scalable recommender system design under reproducible protocols. Source code at \url{https://github.com/sisinflab/A-Reproducible-and-Fair-Evaluation-of-Partition-aware-Collaborative-Filtering}.

\end{abstract}
\section{Introduction}

\noindent\textbf{Motivation.} Over the last decades, recommender systems have evolved through multiple waves of methods, from early neighborhood and latent-factor models to contemporary neural and graph-based approaches \cite{DBLP:journals/tkde/AdomaviciusT05,MF}. Within this broader landscape, two influential strands have emerged: \emph{graph-based} models, which propagate signals over user--item graphs to capture high-order relations \cite{DBLP:conf/iclr/KipfW17,NGCF,LightGCN}, and \emph{item--item similarity} models, which leverage local neighborhoods to deliver simple and effective recommendations \cite{sarwar2001,DBLP:journals/internet/LindenSY03,deshpande2004}. Each strand brings distinct strengths and liabilities: graph-based methods model multi-hop structure but typically rely on message passing and iterative training \cite{LightGCN}, whereas item--item approaches enable lightweight inference and transparent pipelines, yet face quadratic similarity computation as catalogs grow (with sparsity-inducing learners like SLIM mitigating scalability in practice) \cite{SLIM}.
\\
\indent Against this background, \emph{partition-aware} strategies have been proposed to balance accuracy and efficiency by explicitly decomposing the item graph into coherent subgraphs and learning local similarities within each part while preserving limited global context. The Fine-tuning Partition-aware Item Similarities (\textsc{FPSR}) framework exemplifies this direction by coupling partition-level training with a small global component, and \textsc{FPSR+} augments it with hub sets that bridge partitions to enhance robustness under small or imbalanced clusters \cite{FPSRplus}. A complementary \emph{block-aware} line, instantiated by the Block-aware Item Similarity Model (\textsc{BISM}), regularizes a global similarity matrix so that item communities emerge as approximately block-diagonal structures during training \cite{BISM}. Both perspectives are grounded in the empirical regularity that items in large catalogs tend to organize into cohesive communities (e.g., through co-purchase and substitution/complement networks) \cite{DBLP:conf/sigir/McAuleyTSH15,DBLP:journals/tweb/LeskovecAH07}. The main difference is how this prior is designed: through explicit partitioning and hubs in FPSR/FPSR+ versus implicit block structure via regularization in BISM, resulting in distinct computational profiles. Given that these models leverage naturally occurring clustering yet differ fundamentally in their design choices, it is essential to investigate their performance using transparent splits, rigorous hyperparameter optimization, and metrics that capture both accuracy and beyond-accuracy dimensions.\\
\noindent\textbf{Open Issues.} Despite promising reports, several issues motivate a deeper, transparent reassessment of partition-aware models. \emph{Replicability} remains a concern~\cite{DBLP:journals/umuai/BelloginS21,dacrema2019,dacrema2025}: prior evaluations have relied on dataset splits that are not always public and, in some cases, validation procedures that may blur selection boundaries, complicating independent verification. \emph{Fair comparison} is also open~\cite{DBLP:conf/recsys/SunY00Q0G20}: the claimed advantages of FPSR/FPSR+ over strong similarity-based baselines such as BISM warrant re-evaluation on a level playing field with consistent data splits and comparable hyperparameter search across competitors. Both FPSR/FPSR+ and BISM utilize natural clustering in user-item data, but they differ in \textit{key design choices}. These differences raise questions about how each method shares information across clusters and ultimately impacts the quality of recommendations. Moreover, the capability of this family of systems to recommend {\em long-tail} items (i.e., those that receive fewer interactions), a property crucial for real-world applications \cite{jannach2013,ge2010}, has not yet been thoroughly investigated under matched protocols. FPSR/FPSR+ and BISM both leverage naturally occurring clustering, yet they differ in terms of global versus local optimization and the roles of global connectors (hub sets) and global regularization. Therefore, it is essential to systematically investigate their performance using transparent splits and metrics, with a particular focus on their effects on both the head and long tail of the catalog, a well-documented challenge in recommendation practice \cite{park2008}.\\
\noindent\textbf{Our Contributions}. We present a rigorous and reproducible evaluation of partition-aware collaborative filtering with FPSR and FPSR+, contrasted against BISM and strong baselines, following a unified and transparent protocol. Our contributions are fourfold:
\begin{itemize}
  \item[(i)] \textbf{Replicability study.} We re-assess FPSR and FPSR+ under the original settings wherever reproducible, documenting where results align and where discrepancies arise due to unstated or unavailable splits, and clarifying their impact on selection and evaluation.
  \item[(ii)] \textbf{Fair benchmarking.} We present a consistent user-based hold-out splitting strategy and perform extensive, comparable hyperparameter optimization for all methods, including FPSR/FPSR+, BISM, and other state-of-the-art baselines, ensuring a fair basis for comparison.
  \item[(iii)] \textbf{Robustness to partitioning evaluation.} We examine how FPSR and its enhanced variant FPSR+ diverge in key structural design choices, focusing exclusively on architecture-level aspects, explicit item partitioning with specialized hub connector items, the incorporation of global contextual signals, and strategies for robust operation under small or imbalanced clusters. 
  
  \item[(iv)] \textbf{Long-tail analysis.} We perform a fine-grained head/overall/tail breakdown to characterize how FPSR/FPSR+ and BISM treat popular versus niche items, drawing on FPSR+’s hub-selection perspective (e.g., connector vs.\ degree-based hubs) to probe how cross-partition connectivity influences long-tail behavior.
\end{itemize}
This study clarifies the practical gains of partition-aware learning, the competitiveness of block-aware global regularization, and their interaction with beyond-accuracy goals in the long tail, under transparent and reproducible conditions.


\section{Related work and Background}

The earliest collaborative filtering (CF) systems were memory-based neighborhood models (e.g., item-$k$NN \cite{deshpande2004}), which rank candidate items using pre-defined similarity functions over historical co-consumption patterns. These methods are simple and explainable, but they rely on static similarity measures and face a quadratic storage and computation bottleneck as the item catalog grows. By explainable, we mean that recommendations can be justified through a small set of influential neighbor items and their similarity weights, enabling straightforward inspection of the contribution of each neighbor to the final score.
In parallel, model-based approaches emerged, either factorizing the user--item signal into latent embeddings (e.g., MF \cite{MF}) or learning the item--item similarity weights directly as optimization variables (e.g., SLIM \cite{SLIM} and EASE \cite{EASEr}). In addition to these similarity-based models, graph convolutional networks (GCNs) have gained prominence~\cite{LightGCN,UltraGCN,SVD-GCN,NGCF}. GCNs propagate user–item signals through the interaction graph to capture high-order connectivity patterns. Despite their expressive power, GCNs often have high computational overhead and slow iterative training, which limits their scalability in large-scale recommendation scenarios. This shows that data-driven similarity estimation typically surpasses fixed heuristics, while maintaining $k$NN-like efficiency during inference.\\
\indent A consistent empirical observation across learned similarity models \cite{LorSLIM} is that items tend to form \emph{cohesive clusters}, i.e., the learned similarity matrix exhibits an \emph{approximately block-diagonal} structure with strong intra-block and weak inter-block relations. Interestingly, a similar phenomenon can be observed in graph-based models such as GCNs, where sampling or pruning often yields disconnected or weakly connected subgraphs that reflect natural item communities. This suggests that clustering is a unifying structural principle underlying both item-similarity and graph-based paradigms. This emergent clustering behavior had previously been treated as a by-product of learning dynamics rather than an explicit design target. BISM was the first to formalize this phenomenon, introducing it as an architectural prior rather than a side effect.
Its architecture embodies this principle by representing the final item-item similarity matrix $S$ as the sum of two specialized components: a local similarity matrix $S^l$, which captures intra-partition relationships, and a global similarity matrix $S^g$, which models residual connections between the learned clusters.
Starting from a random initialization, BISM adopts a \emph{block-diagonal regularization} (BDR) strategy to force the matrix $S^l$ to have a clustered structure during optimization.
Intuitively, BDR (i) strengthens localized similarity estimation inside clusters, (ii) induces sparsity by dampening cross-cluster edges, and (iii) exploits in-block transitivity to mitigate data sparsity.
However, BISM's training relies on an interdependent alternating optimization of $S^l$ and $S^g$ components. Since both optimization steps operate across the full $N \times N$ item space (where $N$ represents the number of items in the catalogue), the model's overall complexity and memory requirements still scale quadratically with the item catalog size.
\\
\indent To address the scalability bottleneck, FPSR model adopts an explicit \emph{divide-and-conquer} strategy. It first partitions the item–item graph (e.g., via spectral partitioning) into several nearly-disjoint subgraphs. This partitioning is performed recursively and is governed by a size ratio parameter $\tau$, which defines the maximum relative size of any partition, thereby allowing direct control over the granularity of the decomposition. State that, we define $K$ as the final number of partitions computed by this procedure. Each partition represents a distinct subset of the complete item space $N$. For each partition, we define $M_k$ as the count of items contained within that partition. Within these partitions, we calculate a similarity matrix, which captures the relationships and similarities between the items in that subset. This methodology effectively transforms a single, dense optimization problem characterized by $N^2$ dimensions into multiple, more manageable optimization tasks, each with dimensions of $M_k^2$ ($M_k \ll N$). As a result, this partitioning scheme yields substantial improvements in training time and memory efficiency, enabling faster computations and reduced resource consumption. It re-injects global context by constructing a global similarity matrix $W$ from the top eigenvectors of the user-item strategy matrix $S$ via a weighting parameter $\lambda$, forming the final similarity $C = S + \lambda W$. 
FPSR$+$ further strengthens robustness in the presence of small or imbalanced partitions by introducing a \emph{hub set}: a carefully selected subset of cross-partition \textit{bridge} items (e.g., popular ones) whose signals are shared to stabilize local learning and recover inter-partition effects. Intuitively, when partitions become small or imbalanced, local similarity estimation becomes noisy because co-occurrence evidence is fragmented within each subgraph. The hub set acts as a shared set of anchor items across partitions, reintroducing cross-partition signal and reducing variance in the learned similarities, which improves stability under aggressive partitioning. Both FPSR and FPSR$+$ demonstrate competitive accuracy compared to strong baselines while significantly lowering computational costs and substantially reducing the model footprint through partition-wise splitting.
\\
\indent FPSR, as well as BISM, arises from the same structural premise of item clustering, but they operate differently. 
BISM treats the cluster structure as a learned property: a single, global similarity matrix is optimized with a BDR penalty that softly suppresses cross-block edges; the number/shape of blocks is data-driven and governed by regularization strength. 
FPSR hard-codes the structure a priori by partitioning the item graph before learning; similarities are then learned only within partitions, with a small global/spectral term and, in FPSR$+$, with a hub set to transmit essential cross-partition information, ensuring stability and recovering inter-partition signals, particularly for small or imbalanced partitions. 
From a computational perspective, BISM’s optimization remains global (quadratic in $M$), whereas FPSR decomposes the problem into a sum of smaller quadratics (roughly $\sum_K M_k^2$). 

This contrast between BISM's implicit, in-objective regularization and FPSR's explicit, structural pre-processing marks a fundamental divergence in modern similarity modeling. While BISM scales its single, global optimization through soft, block-aware constraints, FPSR achieves scalability by physically decomposing the problem space a priori. Each approach thus presents a distinct profile of computational complexity, architectural assumptions, and potential trade-offs between local and global signal propagation.
\section{Experimental Setup}~\label{setup}

Our experimental procedure has two phases. First, we conduct a replicability study to validate the original findings. We initially ran a series of replicability tests on the datasets used in the original FPSR study: \textit{Amazon-CDs}~\cite{DBLP:conf/sigir/McAuleyTSH15}, \textit{Gowalla}~\cite{DBLP:conf/kdd/ChoML11}, \textit{Yelp2018}~\cite{DBLP:conf/isspit/KronmuellerCHD18}, and \textit{Douban}~\cite{DBLP:conf/cikm/ShiZLYYW15}. Dataset statistics are reported in~\Cref{stats}. We follow as closely as possible the authors' FPSR setup: we used the original data splits and the hyperparameters reported in their public repository\footnote{https://github.com/Joinn99/FPSR}. The situation is more heterogeneous for the two FPSR+ variants. For FPSR+$_D$, the authors released an implementation limited to experiments on only one dataset, which hinders broader reproducibility. However, we have experimentally observed its hyperparameters coincide with those of FPSR, apart from the introduction of an additional fixed-value parameter. In contrast, no implementation was provided for FPSR+$_F$ variant. Based on the methodological details in the original article, we developed our own implementation, enabling us to reproduce the experimental results with a standard implementation effort. We contacted the original authors; however, we did not receive a response, and we are unable to address this lack of details. More details are explained in \Cref{sec:replicability_section}.

\begin{table}[t!]
\setlength{\tabcolsep}{5pt}
\scriptsize
\centering
\caption{Statistics of the datasets.}~\label{stats}
\begin{tabular}{ccccccc}
\toprule
\textbf{Dataset} & \textbf{Users} & \textbf{Items} & \textbf{Interactions} & \textbf{Density} & \textbf{Gini$_U$} & \textbf{Gini$_I$} \\
\midrule
Amazon-CDs & 43169 & 35648 & 777426 & 0.051\% & 0.5104 & 0.4613 \\
Douban & 13024 & 22347 & 792062 & 0.272\% & 0.6360 & 0.6242 \\
Gowalla & 29858 & 40981 & 1027370 & 0.084\% & 0.4666 & 0.4346 \\
Yelp2018 & 31668 & 38048 & 1561406 & 0.130\% & 0.3921 & 0.5130 \\
\bottomrule
\end{tabular}
\end{table}

For the next phase, we adopted a new and consistent data splitting and evaluation protocol for the same four datasets that were previously mentioned. We employed a user-based hold-out splitting strategy for all datasets. For each user, 15\% of their interactions were randomly sampled to create the test set. Subsequently, from the remaining interactions, another 15\% for each user were sampled to form the validation set. The remaining interactions constituted the training set. For these main tests, accuracy was assessed using Recall and nDCG at $K=\{10, 20\}$. All experiments were conducted using Elliot~\cite{elliot}.

We compared FPSR against a set of well-established baselines: Random, MostPop, Item-kNN \cite{deshpande2004}, RP$^3\beta$ \cite{RP3beta}, EASE$^R$ \cite{EASEr}, GF-CF \cite{GFCF}, LightGCN \cite{LightGCN} and BISM. It is worth noting that Wei et al. \cite{FPSR}
do not report details regarding baselines hyperparameter tuning.
Most baselines are established state-of-the-art models, except BISM, whose results we verified using the authors’ public code since its data split was available only for BookCrossing. Although we omit the reproducibility analysis to focus on FPSR and FPSR+, we include BISM due to its strong similarity and limited use in literature.

Following the rigorous evaluation principles advocated by \cite{dacrema2019}, we took particular care to ensure a fair comparison: we performed a hyperparameter optimization for all baselines by running 20 trials based on the  Tree-structured Parzen Estimator (TPE)~\cite{DBLP:conf/icml/BergstraYC13} strategy, with Recall@20 as the validation metric, ensuring that each algorithm could perform at its maximum potential. TPE is a Bayesian optimization method that models promising hyperparameter regions by comparing configurations that yield good vs. poor validation performance, and then samples new trials from the regions expected to improve the objective. This rigorous methodology allows us to robustly validate the contribution of FPSR and its variants, ensuring that any reported improvement constitutes a genuine advancement over a robust set of competitors.


The complete parameters search spaces for all other baselines and all the best founded configurations are documented in our public code repository\footnote{\url{https://github.com/sisinflab/A-Reproducible-and-Fair-Evaluation-of-Partition-aware-Collaborative-Filtering}}. Due to space constraints, we report here few search spaces for the primary subject of our study, FPSR and its variants: the weight $\lambda$, that is the weight of global information with respect to the local information, from \{0.1, 0.2, $\cdots$, 0.5\}; the size ratio parameter $\tau$, chosen from \{0.1, 0.2, $\cdots$, 0.5\}.


\section{Results}

Here, we connect our four contributions with the following research questions:
\begin{description}
\item[RQ1 (Replicability study).] \textit{Under what conditions can the originally reported results of FPSR and FPSR+ be reliably reproduced, and what discrepancies arise when key experimental details cannot be exactly replicated?}
\item[RQ2 (Fair benchmarking).] \textit{How do FPSR and its extension FPSR+ perform relative to BISM and other baseline models when all methods are evaluated under a fair, unified protocol with identical data splits and consistent hyperparameter tuning?}
\item[RQ3 (Robustness to partitioning evaluation).] \textit{What impact do the core structural differences between FPSR and FPSR+, such as the introduction of hub connector items, the injection of cross-partition interaction, and robustness with respect to partition size or balance, have on the theoretical underpinnings and operational behavior of these partition-aware collaborative filtering models?} 
\item[RQ4 (Long-tail Analysis).] \textit{How do partition-aware models differ from the block-aware model BISM in their ability to recommend head items versus long-tail items, and what role do FPSR+'s hub-based mechanisms play in influencing long-tail recommendation performance?}
\end{description}

\subsection{RQ1: Replicability Study}
\label{sec:replicability_section}

\begin{table}[b!]
\scriptsize
\setlength{\tabcolsep}{7pt}
\centering
\caption{Results of the replicability study across FPSR, FPSR+$_D$, and FPSR+$_F$ on four datasets. R@20 and N@20 represent Recall@20 and NDCG@20. Missing entries (–) denote unvailable data splits.}
\label{tab:replicability_all}
\begin{tabular}{l l cccccc}
\toprule
& \multirow{2}{*}{\textbf{Dataset}} 
& \multicolumn{2}{c}{\textbf{Ours}} 
& \multicolumn{2}{c}{\textbf{Original}} 
& \multicolumn{2}{c}{\textbf{Performance Shift}} \\ 
\cmidrule{3-8} 
& & R@20 & N@20 & R@20 & N@20 & R@20 & N@20 \\ 
\cmidrule{1-8}
\multirow{4}{*}{\rotatebox{90}{\tiny{\textbf{FPSR}}}}
            & Amazon-CDs & 0.1557 & 0.0853 & 0.1576 & 0.0896 & $-1.9\cdot10^{-03}$ & $-4.3\cdot10^{-03}$ \\
            & Douban     & 0.1994 & 0.1713 & 0.2095 & 0.1950 & $-1.01\cdot10^{-02}$ & $-2.37\cdot10^{-02}$ \\
            & Gowalla    & 0.1890 & 0.1575 & 0.1884 & 0.1566 & $+6.0\cdot10^{-04}$ & $+9.0\cdot10^{-04}$ \\
            & Yelp2018   & 0.0704 & 0.0587 & 0.0703 & 0.0584 & $+1.0\cdot10^{-04}$ & $+3.0\cdot10^{-04}$ \\
\cmidrule{1-8}
\multirow{4}{*}{\rotatebox{90}{\tiny{\textbf{FPSR+$_D$}}}}
            & Amazon-CDs & 0.1602 & 0.0878 & 0.1622 & 0.0922 & $-2.0\cdot10^{-03}$ & $-4.4\cdot10^{-03}$ \\
            & Douban     & 0.2096 & 0.1785 & 0.2202 & 0.2035 & $-1.1\cdot10^{-02}$ & $-2.5\cdot10^{-02}$ \\
            & Gowalla    & 0.1891 & 0.1575 & 0.1892 & 0.1577 & $-1.0\cdot10^{-04}$ & $-2.0\cdot10^{-04}$ \\
            & ML-20M     & -      & -      & -      & -      & -                   & -                   \\
\cmidrule{1-8}
\multirow{4}{*}{\rotatebox{90}{\tiny{\textbf{FPSR+$_F$}}}}
            & Amazon-CDs & 0.1573 & 0.0861 & 0.1600 & 0.0908 & $-2.7\cdot10^{-03}$ & $-4.7\cdot10^{-03}$ \\
            & Douban     & 0.2028 & 0.1739 & 0.2082 & 0.1943 & $-5.4\cdot10^{-03}$ & $-2.0\cdot10^{-02}$ \\
            & Gowalla    & 0.1892 & 0.1576 & 0.1893 & 0.1578 & $-1.0\cdot10^{-04}$ & $-2.0\cdot10^{-04}$ \\
            & ML-20M     & -      & -      & -      & -      & -                   & -                   \\
\bottomrule
\end{tabular}
\end{table}
We compared in Table \ref{tab:replicability_all} the values published in the original paper with those calculated in our tests, accompanied by their difference. The performances are practically identical on Gowalla and Yelp2018, exhibiting a marginal performance shift on the order of $10^{-4}$ while on Amazon-CDs and especially on Douban more marked discrepancies emerge. For Gowalla and Yelp2018, the correspondence is explained by the fact that the authors published the exact data splittings they used. However, it should be noted that, based on the configuration file provided in the original FPSR repository, it appears that the test split was also used for model selection. We note that this is an inference from the released configuration and may not fully reflect all experimental choices described in the paper. Conversely, for Amazon-CDs and Douban, the absence of published data splits forced us to generate them by executing the authors' original code. This introduces stochasticity in the data splitting, a known obstacle to replicability of the experiments \cite{DataRec}. The considerable deviations we observed from the published results are a direct consequence of this.


This pattern of discrepancies, tied to the availability of the data splits, extends to the two FPSR+ variants as well, and not only. For FPSR+$_D$, only a single-dataset implementation was released, yet the results were reproducible given its near-identical hyperparameters to FPSR. For FPSR+$_F$, no code was available, so we reimplemented it from the paper’s description. It is important to emphasize that the dataset splitting for ML-20M and the corresponding hyperparameter configurations were not made publicly available for either FPSR+ variant. Consequently, the exact data partitioning used in the original study could not be recovered. This omission prevents complete verification of the original results and ultimately hinders a full assessment of the replicability of FPSR+.

\subsection{RQ2: Fair Benchmarking}
\label{accuracy_section}

\begin{table*}[!t]
\scriptsize
\setlength{\tabcolsep}{4pt}
\centering
\caption{Results are presented in terms of Recall@K (R@K) and nDCG@K (N@K). The best outcomes are highlighted in \textbf{bold}, while the second-best are indicated by \underline{underlined}. A $^{*}$ denotes results that have statistical significance below the 0.05 threshold between the best and second-best.}
\label{tab:accuracy}
\begin{tabular}{l
    cccc
    cccc}
\toprule
& \multicolumn{4}{c}{Amazon-CDs}
& \multicolumn{4}{c}{Douban} \\
\cmidrule(r){2-5}\cmidrule(l){6-9}
  & R@10 & N@10 & R@20 & N@20 & R@10 & N@10 & R@20 & N@20 \\
\midrule
Random          & 0.0002 & 0.0001 & 0.0005 & 0.0002 & 0.0005 & 0.0006 & 0.0012 & 0.0008 \\
MostPop         & 0.0067 & 0.0049 & 0.0129 & 0.0069 & 0.0485 & 0.0511 & 0.0708 & 0.0555 \\
\midrule
Item-kNN        & 0.0837 & 0.0619 & 0.1171 & 0.0726 & 0.1263 & 0.1462 & 0.1721 & 0.1526 \\
RP$^3\beta$     & 0.0532 & 0.0376 & 0.0844 & 0.0475 & 0.1132 & 0.1306 & 0.1626 & 0.1402 \\
EASE$^R$        & 0.0828 & 0.0603 & 0.1180 & 0.0713 & 0.1177 & 0.1288 & 0.1671 & 0.1390 \\
\midrule
GF-CF           & 0.0731 & 0.0523 & 0.1109 & 0.0644 & 0.1103 & 0.1198 & 0.1628 & 0.1302 \\
LightGCN        & 0.0866 & 0.0604 & 0.1304 & 0.0743 & 0.1085 & 0.1112 & 0.1633 & 0.1252 \\
\midrule
BISM            & \textbf{0.1032}$^{*}$ & \textbf{0.0772}$^{*}$ & \textbf{0.1435}$^{*}$ & \textbf{0.0900}$^{*}$ & \underline{0.1516} & \underline{0.1793} & \underline{0.2079} & \underline{0.1881} \\
FPSR            & 0.0969 & 0.0603 & 0.1327 & 0.0845 & 0.1479 & 0.1766 & 0.2034 & 0.1854 \\
FPSR+$_D$       & 0.0970 & 0.0727 & \underline{0.1368} & 0.0852 & \textbf{0.1543}$^{*}$ & \textbf{0.1830}$^{*}$ & \textbf{0.2132}$^{*}$ & \textbf{0.1928}$^{*}$ \\
FPSR+$_F$       & \underline{0.0984} & \underline{0.0738} & 0.1357 & \underline{0.0856} & 0.1482 & 0.1745 & 0.2041 & 0.1841 \\
\midrule

& \multicolumn{4}{c}{Gowalla}
& \multicolumn{4}{c}{Yelp2018} \\
\cmidrule(r){2-5}\cmidrule(l){6-9}
  & R@10 & N@10 & R@20 & N@20 & R@10 & N@10 & R@20 & N@20 \\
\midrule
Random          & 0.0002 & 0.0001 & 0.0003 & 0.0002 & 0.0002 & 0.0002 & 0.0004 & 0.0003 \\
MostPop         & 0.0208 & 0.0170 & 0.0306 & 0.0203 & 0.0094 & 0.0092 & 0.0156 & 0.0115 \\
\midrule
Item-kNN        & 0.1237 & 0.1121 & 0.1788 & 0.1305 & 0.0504 & 0.0515 & 0.0822 & 0.0632 \\
RP$^3\beta$     & 0.0946 & 0.0832 & 0.1493 & 0.1018 & 0.0446 & 0.0473 & 0.0761 & 0.0585 \\
EASE$^R$        & 0.1095 & 0.0990 & 0.1618 & 0.1162 & 0.0488 & 0.0499 & 0.0795 & 0.0611 \\
\midrule
GF-CF           & 0.1341 & 0.1197 & 0.1979 & 0.1411 & 0.0594 & 0.0615 & 0.0987 & 0.0759 \\
LightGCN        & 0.1290 & 0.1155 & 0.1887 & 0.1357 & 0.0566 & 0.0576 & 0.0934 & 0.0711 \\
\midrule
BISM            & \textbf{0.1463} & \textbf{0.1344} & \textbf{0.2072} & \textbf{0.1544}$^{*}$ & 0.0542 & 0.0577 & 0.0871 & 0.0693 \\
FPSR            & 0.1436 & 0.1307 & 0.2062 & 0.1515 & 0.0561 & 0.0589 & 0.0906 & 0.0712 \\
FPSR+$_D$       & \underline{0.1460} & \underline{0.1335} & \underline{0.2064} & \underline{0.1532} & \textbf{0.0595}$^{*}$ & \textbf{0.0626}$^{*}$ & \textbf{0.0965}$^{*}$ & \textbf{0.0759}$^{*}$ \\
FPSR+$_F$       & 0.1405 & 0.1281 & 0.2012 & 0.1482 & \underline{0.0581} & \underline{0.0607} & \underline{0.0944} & \underline{0.0739} \\
\bottomrule
\end{tabular}
\end{table*}
After verifying the replicability of the original findings, we conducted a comprehensive benchmark to assess the performance of the FPSR model family within a fair and rigorous evaluation setting. Unlike the initial replicability phase, this benchmark utilizes a consistent user-based hold-out splitting strategy across all datasets (see \Cref{setup}). We performed extensive hyperparameter optimization for all models to ensure each algorithm was competing at its full potential. This methodology allows for a robust validation of the models' contributions against a strong and diverse set of competitors.

The results presented in \Cref{tab:accuracy} lead to several critical insights, including a significant difference from the conclusions drawn in the original paper. Perhaps the most impactful finding of our benchmark is the re-evaluation of BISM's performance relative to the FPSR family. Our results provide valuable insights, indicating a direct reversal of the performance hierarchy observed in the original findings. This shift presents an opportunity for further exploration and understanding of the underlying factors at play. In our experiments, BISM emerges as the top-performing model on the Amazon-CDs dataset (R@10: 0.1032, N@20: 0.0900), decisively outperforming not only the base FPSR model but also its advanced FPSR+ variants. This is in stark contrast to the original publications, where FPSR was shown to hold an advantage over BISM on this same dataset.
We attribute this significant reversal to our commitment to a thorough and equitable hyperparameter optimization process for all competing models. While the original authors understandably focused on tuning their novel architecture, our independent evaluation created a level playing field. BISM, a model with a distinct block-aware regularization scheme, appears to be particularly responsive to meticulous tuning. Our extensive search likely may have revealed a higher performance ceiling for BISM than what was achieved in the original comparisons. This outcome serves as a critical lesson for recommender system evaluation. It suggests that the reported superiority of a new model can sometimes be an artifact of a focused tuning effort rather than an inherent architectural advantage. A strong, established baseline like BISM, when afforded the same level of rigorous optimization, can be an even more powerful competitor \cite{dacrema2019}.
\\
\indent While BISM claimed the top spot in Amazon-CDs, the FPSR family showcased its capabilities by consistently leading performance across the other three datasets. Our rigorous benchmarking presents a clearer view of the evolution from FPSR to FPSR+. Although the FPSR+ variants often achieve the highest performance, the original FPSR model remains impressively strong, with marginal differences in performance in several cases. Notably, the performance of the base FPSR model is nearly identical to that of FPSR+ on Gowalla.
\begin{table*}[!t]
\scriptsize
\setlength{\tabcolsep}{12pt}
\centering
\caption{The table reports the performance of FPSR, FPSR$+_D$, and FPSR$+_F$ under different $\tau$ values on Douban dataset. The value of $\tau_\text{{best}}$ is taken from the best configuration founded in our experimental analysis.}
\label{tab:sensitivity}
\begin{tabular}{l
    cccc
    cc}
\toprule
& \multicolumn{2}{c}{FPSR($\tau_\text{{best}}=0.5$)}
& \multicolumn{2}{c}{FPSR$+_D$($\tau_\text{{best}}=0.4$)}
& \multicolumn{2}{c}{FPSR$+_F$($\tau_\text{{best}}=0.4$)} \\
\cmidrule(r){2-3}\cmidrule(r){4-5}\cmidrule(r){6-7}
  & R@20 & N@20 & R@20 & N@20 & R@20 & N@20 \\
\midrule
$\tau = 0.05$           & 0.1631 & 0.1561 & 0.2068 & 0.1882 & 0.1731 & 0.1602 \\
$\tau = 0.15$           & 0.1918 & 0.1767 & 0.2097 & 0.1908 & 0.1953 & 0.1771 \\
$\tau = 0.25$           & 0.1930 & 0.1780 & 0.2109 & 0.1912 & 0.1957 & 0.1780 \\
\midrule
$\tau_\text{{best}}$    & 0.2034 & 0.1855 & 0.2132 & 0.1928 & 0.2041 & 0.1841 \\
\bottomrule
\end{tabular}
\end{table*}


\subsection{RQ3: Robustness to Partition Granularity and Imbalance}

The previous section has shown no fixed performance hierarchy within the FPSR family; the relative ranking of FPSR, FPSR+$_D$, and FPSR+$_F$ depends on the evaluation setting. Unless otherwise noted, all results in this subsection follow our experimental protocol: we use our own dataset splits and run each method with the best-performing hyperparameters selected on validation (our \textit{best} params). We hypothesize that careful hyperparameter optimization, especially of the partition-size parameter $\tau$, can substantially mitigate the base FPSR’s sensitivity to small or imbalanced partitions. FPSR+ was explicitly designed to address this limitation by introducing hub items that act as cross-partition bridges, preserving inter-partition signal when partitions become too fine-grained or skewed in size.

We therefore analyze how accuracy varies with $\tau$ for all FPSR variants. Table 4 summarizes the trend, in the Douban dataset: when $\tau$ is very small ($\tau$ = 0.05), the base FPSR exhibits a notable degradation because each partition contains few training signals and limited cross-partition context. Concretely, FPSR’s Recall@20 drops from 0.2034 at its best setting ($\tau_\text{best}$ = 0.5) to 0.1631 ($\approx$ 20\% relative decrease), with a similar decline in nDCG@20. In contrast, FPSR+$_D$ remains far more stable ($\approx$ 3\% decrease), indicating that popular hubs effectively buffer the loss of cross-partition information. FPSR+$_F$ also improves robustness over vanilla FPSR, though the stabilization is more modest on this dataset. The nDCG@20 patterns mirror these observations.

These results support two conclusions. First, systematic tuning of $\tau$ narrows the gap between FPSR and its hub-augmented variants when partitions are not overly tight. Second, under aggressive partitioning or pronounced imbalance, hub mechanisms provide resilience by sharing salient signals across partition boundaries, thus stabilizing local learning and retaining inter-partition knowledge. The results for all datasets are available in our repository.

\subsection{RQ4: Long-tail Analysis}

The original FPSR+ paper provides a preliminary analysis of the models' performance on popular versus long-tail items to motivate its architectural choices. It suggests a clear separation of tasks: the degree-based hub selection (FPSR$+_D$) is adopted to enhance accuracy on head items, while the Fielder-based strategy (FPSR$+_F$) is proposed to improve performance in the long tail. This claimed separation directly addresses the fundamental recommender trade-off between exploiting popular items and exploring the catalog. Our comprehensive benchmark (\Cref{accuracy_section}) already established the overall performance hierarchy. Analyzing these results by item popularity, we can understand them more deeply. To this end, we divide the elements into two subsets based on their popularity, i.e., based on the interaction frequency in the training set: following common practice in long-tail analysis, the top 10\% of the most popular items are defined as the head, while the remaining 90\% constitute the tail. Consequently, three evaluation sets are derived: the overall test set includes all items, the head test set contains only head items, and the tail test set includes only tail items. For full transparency, also the data splits used in this analysis are made publicly available in our repository. \Cref{tab:results_longtail} presents this breakdown. While confirming the same top-performing models, this more complex view reveals the distinct strengths and weaknesses each model family exhibits across the popularity spectrum.

\begin{table*}[!t]
\scriptsize
\setlength{\tabcolsep}{9pt}
\centering
\caption{Results in terms of Recall@20 and nDCG@20 on four datasets. Best values are in \textbf{bold}, second-best are \underline{underlined}.}
\begin{tabular}{llcccccc}
\toprule
& \multirow{2}{*}{\textbf{Model}} & \multicolumn{3}{c}{Recall@20} & \multicolumn{3}{c}{nDCG@20} \\
\cmidrule(lr){3-5} \cmidrule(lr){6-8}
& & Head & Overall & Tail & Head & Overall & Tail \\
\midrule

\multirow{4}{*}{\rotatebox{90}{\parbox{0.09\linewidth}{\linespread{1}\selectfont\textbf{\tiny{Amazon-\\CDs}}}}}
& BISM        & \textbf{0.2404} & \textbf{0.1435} & \underline{0.0819} & \textbf{0.1397} & \textbf{0.0900} & \underline{0.0409} \\
& FPSR        & 0.2068 & 0.1327 & \textbf{0.0837} & 0.1200 & 0.0845 & \textbf{0.0448} \\
& FPSR$+_D$  & \underline{0.2337} & 0.1368 & 0.0742 & \underline{0.1318} & \underline{0.0852} & 0.0388 \\
& FPSR$+_F$  & 0.2315 & \underline{0.1371} & 0.0756 & 0.1306 & \underline{0.0852} & 0.0394 \\
\midrule

\multirow{4}{*}{\rotatebox{90}{\textbf{\tiny{Douban}}}} 
& BISM        & \textbf{0.2886} & \underline{0.2079} & 0.0628 & \textbf{0.2250} & \underline{0.1881} & 0.0367 \\
& FPSR        & 0.2749 & 0.2034 & \underline{0.0699} & 0.2123 & 0.1854 & \underline{0.0478} \\
& FPSR$+_D$  & \underline{0.2866} & \textbf{0.2132} & \textbf{0.0751} & \underline{0.2196} & \textbf{0.1928} & \textbf{0.0493} \\
& FPSR$+_F$  & 0.2760 & 0.2041 & 0.0682 & 0.2121 & 0.1841 & 0.0458 \\
\midrule

\multirow{4}{*}{\rotatebox{90}{\textbf{\tiny{Gowalla}}}} 
& BISM        & 0.3622 & \textbf{0.2072} & \underline{0.1081} & \textbf{0.2355} & \textbf{0.1544} & \underline{0.0551} \\
& FPSR        & \textbf{0.3693} & 0.2063 & 0.1013 & \underline{0.2354} & 0.1515 & 0.0510 \\
& FPSR$+_D$  & \underline{0.3624} & \underline{0.2065} & 0.1059 & 0.2329 & \underline{0.1532} & 0.0550 \\
& FPSR$+_F$  & 0.3445 & 0.2061 & \textbf{0.1149} & 0.2209 & 0.1517 & \textbf{0.0601} \\
\midrule

\multirow{4}{*}{\rotatebox{90}{\textbf{\tiny{Yelp2018}}}} 
& BISM        & 0.1668 & 0.0871 & 0.0270 & 0.1016 & 0.0693 & 0.0153 \\
& FPSR        & 0.1755 & 0.0906 & \underline{0.0273} & 0.1049 & 0.0712 & \underline{0.0161} \\
& FPSR$+_D$  & \underline{0.1811} & \textbf{0.0951} & \textbf{0.0306} & \underline{0.1096} & \textbf{0.0748} & \textbf{0.0179} \\
& FPSR$+_F$  & \textbf{0.1930} & \underline{0.0944} & 0.0215 & \textbf{0.1145} & \underline{0.0738} & 0.0125 \\
\bottomrule
\end{tabular}
\label{tab:results_longtail}
\end{table*}


A clear emerging pattern is the systematic superiority of the FPSR family in the long-tail region. Across all four datasets, an FPSR variant consistently outperforms BISM for tail items, often by a significant margin. While BISM excels at leveraging strong local co-occurrence signals, the graph-based nature of FPSR can better aggregate weaker signals and uncover latent relationships. Conversely, the more contested performance on head items appears to be dictated by the statistical properties of the data, rather than by simple heuristics like dataset size. On the contrary, the more complex performance landscape for head items is driven not by simple heuristics like dataset size, because performance appears to be dictated by the statistical properties of the dataset. BISM dominates on domains like Amazon-CDs, where strong co-occurrence structures provide solid local signals for its similarity-based approach, and where dense interaction distribution (high Gini index) may favor models that emphasize popular items. The FPSR family, however, excels on the head of datasets like Yelp2018, where other data characteristics may destroy pure local signals, favoring FPSR's ability to aggregate information globally through its graph structure.


The FPSR+ paper hypothesized a clear division of labor between its two hub-selection strategies: the degree-based hub variant FPSR$+_D$ was intended to enhance recommendation accuracy for popular head items, while the Fiedler eigenvector–based hubs in FPSR$+_F$ were proposed to improve performance on the long-tail. However, our head-to-tail analysis suggests that this separation is not strictly maintained across all datasets. In fact, FPSR$+_F$ sometimes outperforms FPSR$+_D$ on head items (e.g. delivering the best head-item accuracy on Yelp2018) and simultaneously achieves the strongest tail performance on other data (e.g. the best long-tail results on Gowalla). Hence, the effectiveness of each hub selection strategy is dataset-dependent rather than fixed: the supposed head-vs-tail specialization can flip depending on the characteristics of the dataset. Nevertheless, a consistent benefit observed with both FPSR$+$ variants is the mitigation of long-tail performance degradation that was seen in the vanilla FPSR. The inclusion of hub items prevents the sharp drop-off in tail-item accuracy that the base partitioned model can experience, especially in scenarios with very small partitions or sparse training signals. By sharing key signals through global hubs, FPSR$+_D$ and FPSR$+_F$ bolster the recommendation quality for less-popular items, alleviating the tail weakness of the original FPSR framework. This use of hubs to stabilize tail performance was a central motivation for FPSR$+$ and is borne out in our results, with the FPSR family consistently avoiding the severe tail-item losses observed when no hub mechanism is in place.

\section{Conclusion}

In our replication study, we observed that the original FPSR results are only partially reproducible. We could match the reported performance using the official data splits, but missing implementation details prevented complete verification of the FPSR+ method. This motivated us to create a benchmark for re-evaluation.

In this evaluation, we observed that the block-aware BISM model can outperform the FPSR model when both are fairly optimized, contradicting the original findings. BISM achieved higher accuracy than FPSR (including FPSR+) on datasets like Amazon-CDs and Gowalla. However, FPSR remains competitive, achieving state-of-the-art results on other benchmarks, with only a small performance gap between the base FPSR and its enhanced variants.
The performance gap between the base FPSR model and its enhanced variants is notable. Adding global hub items in FPSR+ improves robustness and reduces long-tail item accuracy drops, especially with small partitions or imbalanced data. However, the effectiveness of hub-selection strategies varies by dataset: FPSR$+_D$ typically excels with popular items, while FPSR$+_F$ is expected to perform better with long-tail items. Nevertheless, this is not always the case, as FPSR$+_F$ sometimes surpasses FPSR$+_D$ in head-item accuracy and vice versa for tail items.

To summarize our findings, we highlight the trade-offs between model structure, robustness, and generalization. While partition-based methods with global hubs improve robustness for niche long-tail content, simpler methods may generalize better in some cases. Thus, no single approach is superior across all metrics. Future research should explore alternative hub-selection strategies, integrate graph neural architectures like GCNs, and validate these methods' effectiveness on large-scale industrial datasets.


\section*{Ack}
This work was partially supported by the following projects: P+ARTS - Partnership for Artistic Research in Technology and Sustainability, funded by the European Union Next- GenerationEU (NRRP – M4C1, Investment 3.4, INTAFAM00037, CUP: G43C24000640006).

\bibliographystyle{splncs04}
\bibliography{our}

\end{document}